\documentclass[twocolumn,prl,aps]{revtex4}
\usepackage{amsmath}
\usepackage{amssymb}
\usepackage{graphicx}
\usepackage{dcolumn}
\usepackage{bm}
\usepackage{xcolor}
\usepackage{epstopdf}

\begin{document}

\preprint{Phys.Rev.Research}

\title{Magnetohydrodynamics and electro-electron interaction of massless Dirac fermions.}
\author{D. A. Khudaiberdiev,$^{1,3}$ G. M. Gusev,$^2$ E. B. Olshanetsky, $^{1}$
 Z. D. Kvon,$^{1,3}$ and N. N. Mikhailov $^{1,3}$ }

\affiliation{$^1$Institute of Semiconductor Physics, Novosibirsk
630090, Russia}
\affiliation{$^2$Instituto de F\'{\i}sica da Universidade de S\~ao
Paulo, 135960-170, S\~ao Paulo, SP, Brazil}

\affiliation{$^3$Novosibirsk State University, Novosibirsk 630090,
Russia}

\date{\today}
\begin{abstract}
The magnetotransport properties of massless Dirac fermions in a gapless HgTe quantum well are investigated. In samples with narrow channels, a large negative magnetoresistance with a Lorentzian profile is observed, which is interpreted as a manifestation of electron viscosity due to electron-electron interaction. Comparison of experiment with theory yields the shear stress relaxation time of the Dirac fermions caused by electron-electron scattering.

\end{abstract}

\maketitle

In many cases, for the description of phenomena in solid-state structures, the approximation of non-interacting electrons turns out to be inapplicable and the electron-electron (e-e) interaction, which is one of the oldest problems in solid state physics, acquires a key role \cite{quinn, chaplik}. The reduction of  dimensionality (D) from 3D to 2D sharply enhances the importance of e-e interaction effects \cite {quinn2}. Moreover, it has been discovered recently that relaxation of perturbations of various types in 2D Fermi system, could be both quantitatively and qualitatively different from the quasiparticle lifetime \cite{principi, narozhny, alekseev}.

In the so called hydrodynamic regime of electron transport the shear stress relaxation time $\tau_{2,ee}$  is responsible for shear viscosity which describes the friction between adjacent layers of liquid moving with different velocities. The index 2 in the e-e scattering time subscript $\tau_{2,ee}$ means that the viscosity coefficient is determined by the relaxation of the second harmonic of the distribution function \cite{alekseev}. The hydrodynamic regime requires $l/l_{2,ee} \gg  1$ and  $l_{2,ee}/W\ll 1$, where $l=v_{F}\tau$ is the electron transport mean free path related to momentum relaxation time ($\tau$) brought about by scattering on defects and phonons, $v_{F}$ is the Fermi velocity,  $W$ is the  channel  width, and $l_{2,ee}=v_{F}\tau_{2,ee}$ is the mean  free path for shear viscosity relaxation \cite{gurzhi}-\cite{raichev}. In the presence of the perpendicular magnetic field B the shear viscosity becomes a tensor depending on B, which leads to giant negative magnetoresistance with a Lorentzian profile in narrow channel devices \cite{alekseev1}. It is possible to extract shear viscosity and shear stress relaxation time from the comparison of the experimental data and the theory. This has been preformed in Refs. \cite{gusev1, gusev2, levin, gusev3} for high-mobility GaAs quantum wells with parabolic spectrum. The study of the temperature dependence of viscosity allows one to distinguish between the electron hydrodynamics and ballistic transport \cite{raichev} and explain the difference between e-e scattering in a single well and in a bilayer \cite{gusev4}.

\begin{figure}[ht!]
\includegraphics[width=8cm]{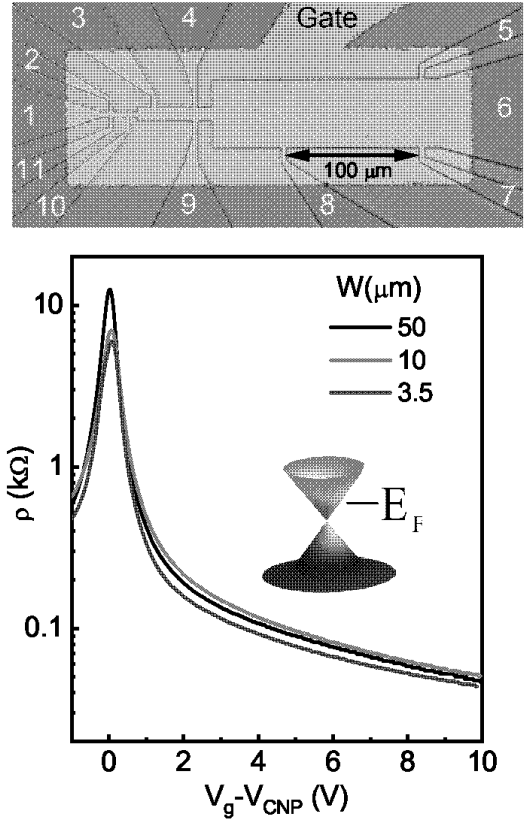}
\caption{(Color online) Schematic of the transistor and the top view of the sample. Resistivity $\rho$ as a function of gate voltage measured for sample sections of different size, $T = 4.2$ K.}
\end{figure}

\begin{figure}[ht!]
\includegraphics[width=7cm]{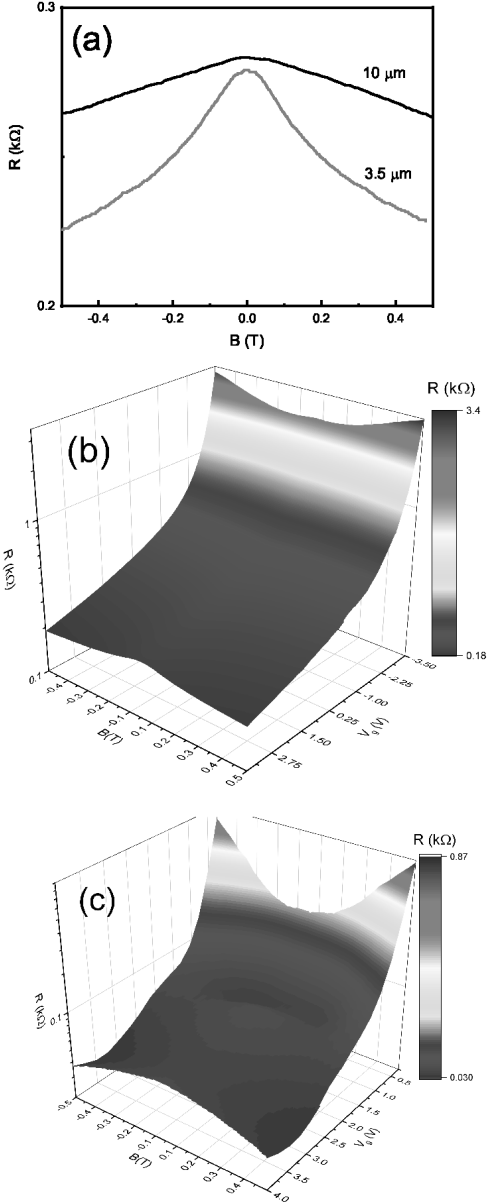}
\caption{(Color online) The resistance R(B) as a function of magnetic field  at $V_{g}=4$ V for sample sections of different size (a) and resistance as a function of gate voltage and magnetic field for the sample section widths $W=3.5\:\text{$\mu$m}$ (b) and  $W=10\:\text{$\mu$m}$ (c), $T=4.2$ K.}
\end{figure}

The hydrodynamic approach is valid for systems with a linear spectrum, such as graphene. Moreover, the electron hydrodynamics in graphene has aroused big interest \cite{polini} because the observation of the negative nonlocal resistance due to current whirpools  \cite{bandurin1} and the superballistic flow in graphene point contacts \cite{kumar} open up the opportunity to study the so called Dirac liquids. The shear stress relaxation time calculation in Refs. \cite{principi, narozhny} highlights the subtle effect of the linear dispersion that distinguishing graphene from the usual parabolic semiconductor-based 2D systems.

Contrary to the quasiparticle lifetime in two-dimensional system which one expects to be expressed as $\tau_{ee}\sim \frac{\hbar E_{F}}{[(kT)^{2}ln(E_{F}/kT)]}$, it has been predicted  that $\tau_{2,ee}\sim \frac{\hbar E_{F}}{(kT)^{2}}$ for a Fermi gas and $\tau_{2,ee}\sim \frac{\hbar ln^{2}(E_{F}/kT)}{(kT)^{2}}$ for a strongly interacting Fermi liquid \cite{alekseev, novikov}. Proportionality coefficients depend on the dispersion relations and are different for graphene and conventional 2D systems \cite{principi,narozhny,alekseev}.

It has been demonstrated that, similar to single layer graphene, a gapless phase of single valley Dirac fermions exists in symmetric HgTe quantum wells with the critical width $d_{c}=6.3$\:nm \cite{buttner, kozlov}. The equation describing the electron linear dispersion relation near Dirac point is $ E(k)=\hbar v_{F}k$, where Fermi velocity $v_{F}=7\times10^{7}\text{\:cm/s}=c/430$ ($c$ is light velocity) is close to the Fermi velocity in graphene $v_{F}=c/300$. Various methods have been proposed for observation of viscosity in the hydrodynamic regime in graphene \cite{bandurin1,bandurin2}. Although experimental studies of transport in a magnetic field are mainly focused on observing the Hall viscosity effect \cite{berdyugin, kim}, subsequent analysis of the results obtained may provide more information on the mechanism of electron-electron shear stress scattering of Dirac fermions.

\begin{figure}[ht!]
\includegraphics[width=9cm]{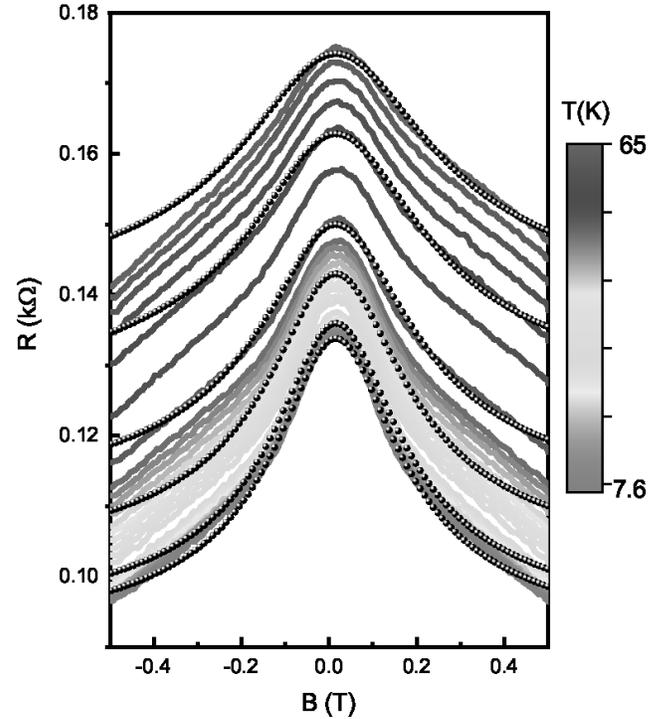}
\caption{(Color online) The resistance  R(B) as a function of magnetic field for different temperatures in a sample segment with a width of 3.5 \text{$\mu$}m for $V_{g}=13$ V.
The circles show the resistance R(B) calculated from
 Eq. (1) for different temperatures T(K): 7.6, 19.2, 38.5, 46.1, 56.6, 64.}
\end{figure}

This paper presents the results of an experimental study of the magnetotransport properties of two-dimensional Dirac fermions in gapless HgTe quantum wells in a wide temperature range. In samples with a small channel width at a high electron concentration, a large negative magnetoresistance with a Lorentzian profile was found, which is in agreement with the hydrodynamic model. Comparing
experiment with theory, we were able to find
the stress relaxation time in a system with a linear spectrum.

\begin{figure}[ht!]
\includegraphics[width=9cm]{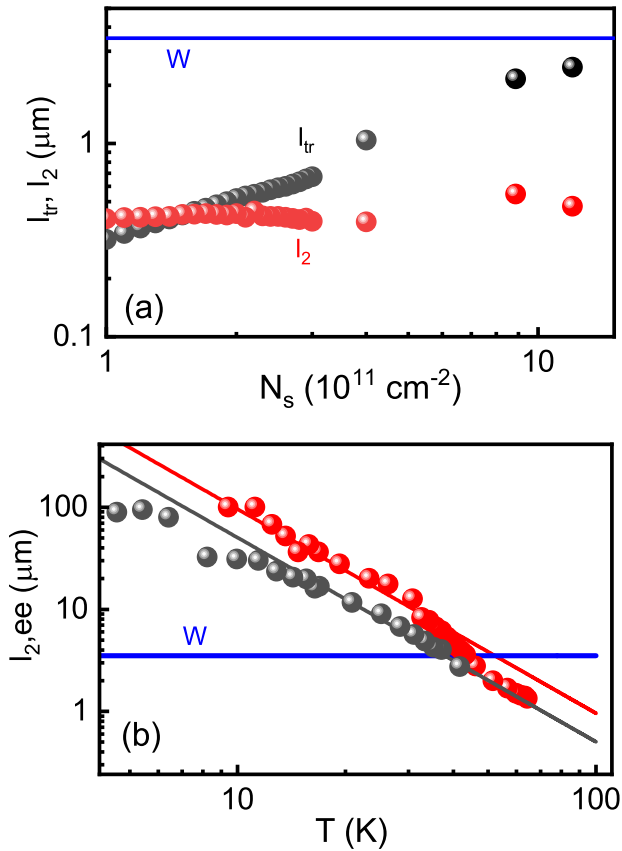}
\caption{(Color online)
(a) The relaxations lengths $l_{tr}$ and $l_{2}$ as a function of the densities, $T=4.2 K$.
(b)The relaxations length  $l_{2,ee}$ as a function of the temperature for two values of the Fermi energy $E_{F}=91$ meV ( black points) and $E_{F}=114$ meV (red points). The thick red and black lines
represent theory with parameters indicated in the  text. Blue line is the channel width.}
\end{figure}

Quantum wells $Cd_{0.65}Hg_{0.35}Te/HgTe/Cd_{0.65}Hg_{0.35}Te$ with (013) surface orientations and a well thickness of 6.3\:nm were prepared by molecular beam epitaxy \cite{kozlov}.
The experimental structures were identical Hall bar devices consisting of three consecutive sections with different widths W: 3.5, 10 and 50\:\text{$\mu$m}, respectively.
The distance between the voltage probes was $L=12, 30, 100\text{\:$\mu$m}$, (fig.1). In addition, a macroscopic Hall bar device with eight voltage probes was also examined. This Hall bar had the width $W$ of $50\text{\:$\mu$m}$ and three consecutive segments of different length $L$ $(100, 250, 100\text{\:$\mu$m})$ (not shown).
A dielectric layer (200\:nm of $SiO_{2}$) was deposited on the sample surface and then covered by a TiAu gate.
The density variation with gate voltage was $1.1\times10^{11}$\text{\:cm$^{-2}$V$^{-1}$}. The resistance in the presence of the perpendicular
magnetic field, R(B), has been measured in the temperature range 4.2 - 70\:K  using a standard four point circuit with a 1-13\:Hz ac current of 1-10\:nA through the sample.

Figure 1 presents the resistivity $\rho= R \frac{W}{L}$ as a function of gate voltage measured at zero magnetic field for segments of different width. One can see that the resistivity is weakly size dependent and demonstrates a typical peak near the charge neutrality point with a value close to $h/2e^{2}$. Further we focus on the sample behavior in the positive gate voltage region, corresponding to the Fermi level residing in the conduction band. In contrast to graphene, the density of states in a zero gap HgTe QW is not symmetric and it rapidly grows in the valence band when moving away from the Dirac point\cite{kozlov}.

Figure 2(a) shows  the resistance as a function of magnetic field  for two narrow segments of the mesoscopic sample with the widths 3.5 and 10\:$\text{$\mu$m}$ at $V_{g}=4\text{\:V}$.
One can see a large negative magnetoresistance ( $R(B)-R(0) < 0)$) with a Lorentzian profile for the 3.5\:$\text{$\mu$m}$ segment and a wider triangular-shaped peak for the 10\:$\text{$\mu$m}$ segment.
The magnetoresistance for the macroscopic sample is shown in the supplementary materials \cite{suppl}.
It is positive for all gate voltages.
The evolution of R(B) with gate voltages in mesoscoscopic samples is given in Figures 2 b and c. One can see that in the vicinity of the CNP the magnetoresistance becomes positive.
Figure 3 shows the representative traces illustrating the evolution of R(B) with temperature for a fixed gate voltage far away from the CNP $V_{g}=13\text{\:V}$.
The resistance curves is almost independent of $T$ at temperatures below 15\:K, but R(B) becomes wider at high temperatures.
Note that the resistance at $B=0$ increases with $T$.
It was found that the $R(T)$ dependences above 4.2\:K can be well described by a cubic law: $R(T)/R(4.2) = 1+ \alpha_{mes} T^{3}$, with $\alpha_{mes} =1.2\times10^{-6}\:$\text{K$^{-3}$}. In the macroscopic sample the $R(B=0)$ above 4.2 K can be approximated with only a quadratic term: $R(T)/R(4.2) = 1+ \alpha_{macr} T^{2}$, with $\alpha_{macr} =8.6\times10^{-5}\:$\text{K$^{-2}$} \cite{suppl}.
The temperature independent term is described by the interface roughness scattering \cite{dobretsova}, while the temperature dependent contribution to the scattering is expected to be due to the phonons, similar to 2D GaAs systems \cite{gusev1}.
Note, however, that the scattering by acoustic phonons leads to a linear rather than a $T^{2}$ dependence for $T > T_{BG}\sim 4 K $, where $T_{BG}$ is Bloch-Gruneisen temperature\cite{sizov}. Further theoretical study is required for the explanation of this behaviour, which is out of scope of this experimental paper.

In existing theories, electron transport in mesoscopic samples is considered within the framework of ballistic, hydrodynamic, or more general models, all based  on a detailed approach assuming solution of the  Boltzmann  kinetic  equation  complemented  with  the  boundary  conditions  for  the  electron distribution function \cite{gurzhi, torre, alekseev1, lucas, scaffidi, lucas2, pellegrino2, holder, raichev}.
The hydrodynamic description of transport in graphene and other Dirac materials have been mostly focused on the vicinity of the Dirac point, where several anomalies and collective excitations in the Dirac fluid have been predicted \cite{briskot, svintsov, molenkamp2, fong}.
Far away from the Dirac point, the system  is expected to be similar to an ordinary Fermi liquid \cite{fong,polini}.

Below we make use of the model proposed in \cite{alekseev, scaffidi, alekseev1}, because it captures all major magnetohydrodynamic properties, including the subtle effects related to the relaxation of the second harmonic of the distribution function by defects and e-e scattering.

The model describes the conductivity as a sum of two independent contributions: the first one is determined by ballistic effects or static disorder and the second one is due to viscosity \cite{alekseev1}. This approach assumes the use of the magnetic field dependent viscosity tensor and the derivation of the resistivity tensor \cite{alekseev1}:

\begin{equation}
\rho(B)= \rho_{0}\left(1+\frac{\tau}{\tau^{*}}\frac{1}{1+(2\omega_{c}\tau_{2})^{2}}\right),\,\,\,
%1
\end{equation}

where $\rho_{0}=\frac{m}{e^{2}n\tau}$,  $1/\tau$ is the scattering  rate due to static disorder, $m$ and $n$ are the effective mass and the density, $\tau^{*}=\frac{W(W+6l_{s})}{12\eta}$, where $\eta=\frac{1}{4}v_{F}^{2}\tau_{2}$ is the viscosity. The shear viscosity relaxation rate is given by $\frac{1}{\tau_{2}(T)}=\frac{1}{\tau_{2,ee}}+\frac{1}{\tau_{2,imp}}$. The relaxation rate $\frac{1}{\tau_{2,imp}(T)}$, determined by process responsible for relaxation of the second harmonic of the distribution function, such as scattering by static defects, gives rise to viscosity, while $\frac{1}{\tau_{2,ee}(T)}$ refers to the shear viscosity relaxation due to e-e scattering \cite{alekseev, alekseev1}. The momentum relaxation rate is expressed as $\frac{1}{\tau}=\frac{1}{\tau_{0, ph}}+\frac{1}{\tau_{0, imp}}$, where $\tau_{0, ph}$ is the term responsible for phonon scattering, and $\tau_{0, imp}$ is the scattering time due to static disorder (not related to the second moment relaxation time) \cite{alekseev1}. The boundary conditions can be characterized by a diffusive scattering or by a slip length $l_{s}$ with extreme cases being no-slip ($l_{s} \rightarrow 0$) and no-stress ($l_{s} \rightarrow \infty$) conditions. It is expected that for $l_{s} \rightarrow \infty$ no hydrodynamic Poiseuille-like flow should be observed.

We fit the magnetoresistance curves and the $R(T)$ at zero magnetic field in Fig. 3 with the 3 fitting parameters : $\tau(T)$, $\tau^{*}(T)$ and $\tau_{2}(T)$. Note that the 2D resistivity is proportional to the resistance: $\rho=\frac{W}{L}R$, and below we discuss the resistivity behaviour. Figure 3 shows the examples of the calculated Lorentzian-like magnetoresistance at different temperatures.  Our  experimental  data  are  in good agreement with the theoretical models \cite{alekseev, alekseev1}. We have also compared magnetoresistance traces for different gate voltages and extracted the relaxation times versus $N_{s}$ dependences. The corresponding relaxations lengths $l=v_{F}\tau$ and $l_{2}=v_{F}\tau_{2}$ are shown in figure 4a. Note, that for massless Dirac fermions the Fermi velocity does not depend on the carrier density. One can see that hydrodynamic conditions are approximately met at high densities, where  we obtain $l/W \sim  1$ and  $l_{2}/W\ll 1$. It is worth noting, that the model \cite{alekseev} is valid for  $l/W \sim  1$, because two  channels providing the relaxation of the second harmonic of the electron distribution function and momentum can be regarded as parallel channels. The transport features associated with the hydrodynamic properties should be observed for $l_{2}/W\ll 1$ and $l_{2}/l\ll 1$, which are reasonably satisfied in our devices at $N_{s} > 4.5\times10^{11} cm^{-2}$ ( Figure 4a).
Note, however, that at densities $N_{s} < 4.5\times 10^{12} cm ^{-2}$, the hydrodynamic conditions are no longer well satisfied in which case both $l_{2}$ and its density dependence become not very meaningful. More stringent conditions $l/W \gg  1$ and  $l_{2, ee}/W\ll 1$ lead to  pure hydrodynamic regime, where second term in the equation 1 becomes dominant and $\rho\approx\frac{m}{e^{2}n}\eta \frac{12}{W^{2}}$ \cite{gurzhi}. In this regime the resistivity is inversely proportional to the square of the temperature, $\rho \sim T^{-2}$, the so-called “Gurzhi effect”, which has been observed in a high quality GaAs 2D electron system \cite{dejong, gusev1, gusev4} and in a graphene quantum  point contact \cite{kim, polini}. Because of a lower mobility, in our HgTe zero gap quantum well the first term in equation 1 will dominate, and the Gurzhi effect is suppressed. Another key parameter that determines the hydrodynamic condition is the slip length  $l_{s}$, which should be much less than the channel width: $l_{s}/W \ll 1$. From comparison with the experiment we derive $\tau^{*}$, and the product  of $\tau^{*} \tau_{2}=W(W+6l_{s})/3v_{F}^2$ yields the slip length $l_{s}$. $l_{s}$ turns out to be negligibly small, confirming the validity of the hydrodynamic description. In addition, this makes it possible to exclude one fitting parameter $\tau^{*}(T)$ from comparison with the theory, which improves the model readability.

Let us now consider the data on the electron-electron interaction that can be obtained from the processing of magnetoresistance. The interparticle scattering time can be found from the equation:

\begin{equation}
\frac{1}{\tau_{2,ee}(T)}=\frac{1}{\tau_{2}(T)}-\frac{1}{\tau_{2,imp}}=C\frac{(kT)^{2}}{\hbar E_{F}}, \,\,\
%2
\end{equation}

where the numerical factor C would be different for systems with parabolic and linear spectrum. For a weakly interacting 2D Fermi gas with a parabolic spectrum $C_{par}$ is defined as \cite{alekseev, novikov}:
$C_{par}=8.4r_{s}^{2}ln\left(\frac{1}{kT/E_{F}+r_{s}}\right)$,
where the interparticle interaction parameter $r_{s}=1/(a_{B}\sqrt{\pi n})$ is small in a Fermi system, $a_{B}$ is the Bohr radius. For a system of massless Dirac fermions parameter $C_{lin}$ takes the form \cite{narozhny}:
$C_{lin}\approx16.4\alpha_{ee}^{2}(ln(E_{F}/kT))$,
where the fine structure constant $\alpha_{ee}$ is defined as $\alpha_{ee}=\frac{e^{2}}{\varepsilon \hbar v_{F}}$ and $\varepsilon$ is the dielectric constant. In Dirac systems the dimensionless parameter and the fine structure constant have the same value. In HgTe QW we find $\alpha_{ee}= r_{s}=3.2/\varepsilon$.

Comparing the temperature dependence of the relaxation rate $1/\tau_{2,ee}(T)$ with Eq.(2), we can derive a temperature-independent characteristic time $\tau_{2,imp}=0.65\times10^{-12}\:\text{s}$. The hydrodynamic approach is associated with a substantial relaxation of the m-th harmonic of the distribution function due to disorder scattering with the rates $\tau_{m,imp}$ \cite{alekseev}. It is usually argued that  $\tau=\tau_{1,imp}$ \cite{alekseev, raichev} and  $\tau_{2,imp}$  are of the same order of magnitude. In our sample we obtain  $\tau_{1,imp}=4.2\times10^{-12}\:\text{s} > \tau_{2,imp}$. Note that in high-mobility GaAs quantum wells it has been found that the time $\tau=\tau_{1,imp}$ is much longer than $\tau_{2,imp}$ (by 10-100 times )\cite{gusev1, gusev4, alekseev}.

Figure 4b shows the  mean free path $l_{2,ee}=v_{F}\tau_{2,ee}$ versus T in a broad temperature range for two values of the Fermi energy $E_{F}=91\:\text{meV}$ and $E_{F}=114\:\text{meV}$. Comparing our results with equation (2) we find parameters $C=1.1$ and $C=0.7$ for lower and higher energies. Thus, comparison with theory turned out to be possible when the temperature changes by one order of magnitude. These values are very different from the values obtained for a 2D electron Fermi system in GaAs well, which are in order of 5-7 \cite{gusev1, raichev, gusev4, alekseev}. Parameters C obtained from the experiment coincide in order of magnitude with the results of calculation, if we assume  the dimensionless  parameter $r_{s}\sim 0.1-0.15$ which is very different from  graphene ($r_{s}\approx 0.7$) due to high dielectric constant.
Although the equations for $C_{par}$ and $C_{lin}$  look different due to logarithmic term, the differences would not be enough to be significant in the temperature dependence.
Discrimination of the subtle difference between massive a massless spectrum is still challenging task and requires more  experimental and theoretical work.

In summary, we have performed a detailed study of the magnetotransport in a single cone massless Dirac fermion system. The negative magnetoresistance was fitted by a Lorentzian profile in accordance with the hydrodynamic approach and the shear stress relaxation time was extracted, which determines the viscosity in the Fermi liquid with a linear dispersion. Compared to graphene, the most typical Dirac material, our system has a number of advantages. First, the fact that only one valley is present in the spectrum allows for a more unambiguous interpretation of the transport measurements, whereas the intervalley scattering present in graphene can affect boundary scattering. Secondly, the advantage of a simpler manufacturing method (MBE growth versus exfoliation) allows for the fabrication of samples of macroscopic size and, therefore, for discriminating between the transport properties due Poiseuille flow in narrow channels and those due to scattering in the bulk.

The financial support of this work by Ministry of Science and Higher Education of the Russian Federation, Grant No. 075-15-2020-
797 (13.1902.21.0024);  São Paulo Research Foundation (FAPESP) Grant No. 2015/16191-5, and the National Council for Scientific and Technological Development (CNPq) is acknowledged. We  thank  O.E. Raichev  for  the  helpful  discussions.

\maketitle
\section{Supplementary material:Magnetohydrodynamics and electro-electron interaction of massless Dirac fermions}

In the main text we have concentrated on the experimental results for samples with the width of 3.5  $\mu m$. At the same time, measurements were also
carried out on a macroscopic Hall bar sample with 8 voltage probes, the width $W=50 \mu m$ and three consecutive segments of different lengths $L$ $(100,
250, 100 \mu m )$ (not shown).

\begin{figure}[ht!]
\includegraphics[width=9cm]{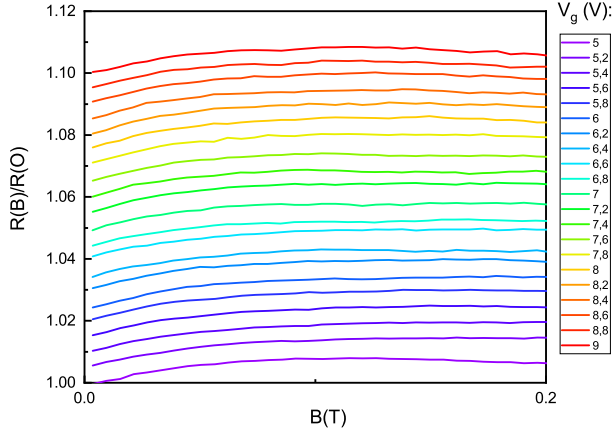}
\caption{(Color online) Normilized resistance  R(B)/R(O) at different gate voltages in the macroscopic sample ($W= 50 \mu m$), T=4.2 K.}
\end{figure}

Figure 1 shows the typical R(B)/R(O) curves for macroscopic sample at different gate voltage in the electron part of the spectrum (the Fermi level
lies in the conduction band). One can see that the magnetoresistance ( $R(B)-R(0) > 0)$) is positive at low magnetic field $B < 0.1 T$, but becomes slightly negative at higher
fields. AT $B > 0.2 T$ we observe Shubnikov- de Haas (SdH) oscillations. In narrow channel samples the magnetoresistance was always negative ( $R(B)-R(0) < 0)$), had a large
$(\sim 50 \%)$ magnitude and a Lorentzian profile, with  SdH oscillations appearing at higher fields. Therefore, one can see that the magnetoresistive is
quite sensitive to the width of the device channel. This observation confirms the assumption about the hydrodynamic origin of the magnetoresistance in
narrow samples.
\begin{figure}[ht!]
\includegraphics[width=9cm]{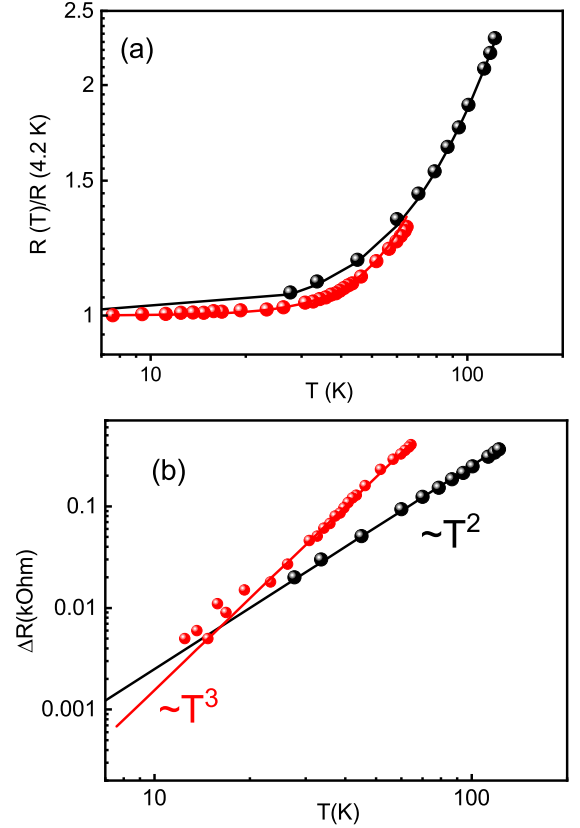}
\caption{(Color online) (a) The normalized resistance as a function of the temperature for the macroscopic and for 3.5  $\mu m$ segment of HgTe QW mesoscopic sample in zero magnetic field. The circles indicate
resistance calculated from Eq. (1). (b) The resistance difference $\Delta R (T)=R(T)-R_{4.2 K}$ as a function of the temperature for the macroscopic (black) and for 3.5  $\mu m$ wide samples (red). Solid lines represents dependencies $\Delta R(T)~ T^{3}$ (red) and $T^{2}$ (black). }
 \end{figure}

As we indicated in the main text the resistance is almost independent of T at temperatures below 15 K and increases with T at higher temperatures. Figure 2 shows the data for macroscopic and mesoscopic sample with the width of $3.5 \mu m$.  It was found that the R(T) dependences in mesoscopic device above 4.2 K can be well described by a cubic law: $R(T)/R(4.2) = 1+ \alpha_{mes} T^{3}$, with $\alpha_{mes} =1.2\times10^{-6} K^{-3}$. In the macroscopic sample the R(B=0) above 4.2 K can be approximated with only a quadratic term: $R(T)/R(4.2) = 1+ \alpha_{macr} T^{2}$, with $\alpha_{macr} =8.6\times10^{-5} K^{-2}$ ( figure 2a).
The figure 2b shows the temperature-dependent difference $\Delta R (T)=R(T)-R_{4.2 K}$ for the macroscopic and mesoscopic samples in log scale.

To further confirm the role of the hydrodynamic factor, we analyze the magnetoresistance in the sample with the width of 10 $\mu m$.
\begin{figure}[ht!]
\includegraphics[width=9cm]{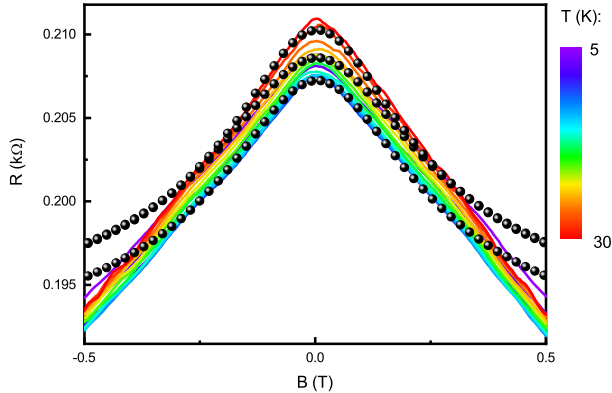}
\caption{(Color online) The resistance  R(B) as a function of the magnetic field in the 10 $\mu m$ segment of HgTe QW sample at $V_{g}=6 V$. The circles indicate
resistance calculated from Eq. (1) in the main text for different temperatures T(K): 5, 14.1, 30}.
 \end{figure}

  \begin{figure}[ht!]
\includegraphics[width=9cm]{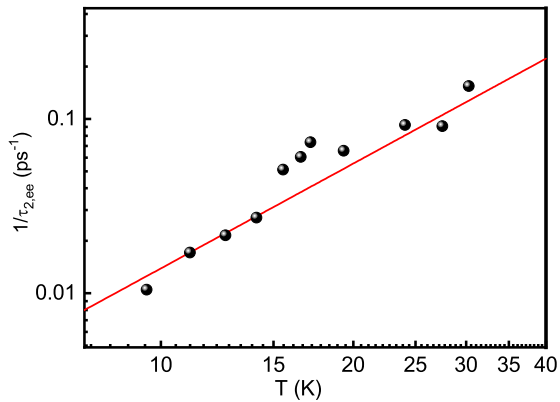}
\caption{(Color online)
The inverse shear relaxation time for the 10 $\mu m$ sample as a function of temperature for the Fermi energy $E_{F}=80 meV$ ( black points). The thick red line
represents the theory with the parameters indicated in the  text.}
\end{figure}
Figure 3 illustrates the variation of R(B) with temperature at a fixed gate voltage $V_{g}=6 V$ away from the CNP. The R(B) is
almost independent of T at temperatures below 15 K, but grows wider at high temperatures. One can see that the curves profile is not exactly Lorentzian,
but rather triangular. It could be assumed that the observed profile distortion is a result of the fact that the hydrodynamic regime is not completely
fulfilled in the wider sample. However, we believe that the model \cite{alekseev} is still valid and the information on the e-e scattering can be extracted
from the data.

The R(B) curves shown in Figure 3 were fitted with the equation (1) of the main text with two adjustable parameters $\tau(T)$, $\tau^{*}(T)$ and
$\tau_{2}(T)$. Note that we have failed in our attempt to improve agreement with theory by extending the range of the magnetic field.  By an appropriate
choice of the fitting parameters $\tau(T)$ and $\tau_{2}(T)$ we obtain $l_{s}=0$ and $W=9.1 \mu m$, which is in excellent agrement with the geometrical
width $W=10 \mu m$. The observed scaling of R(B) with the width strongly supports the validity of the hydrodynamic theory in our samples.
The Poiseuille velocity profile becomes flat with a further increase of the channel width, and the hydrodynamic model becomes inapplicable.

\begin{figure}[ht!]
\includegraphics[width=9cm]{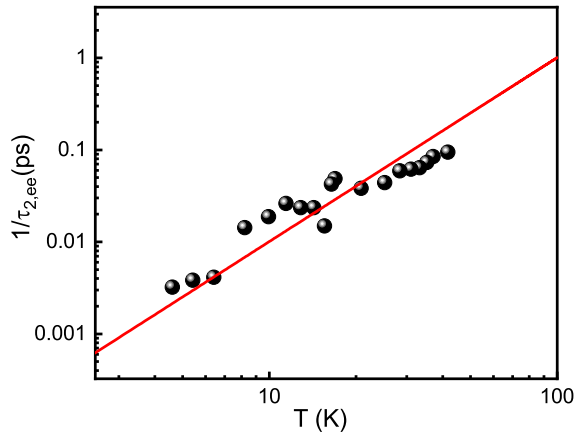}
\caption{(Color online)
The inverse shear relaxation time for the 3.5 $\mu m$ sample as a function of temperature for the Fermi energy $E_{F}=71 meV$ ( black points). The thick red line
represents the theory with the parameters indicated in the  text.}
\end{figure}

Similar to the $3.5 \mu m$ sample, the results for e-e collisions determined from the magnetoresistance data have been compared with the theory:
$\frac{1}{\tau_{2,ee}(T)}=\frac{1}{\tau_{2}(T)}-\frac{1}{\tau_{2,imp}}=C\frac{\hbar E_{F}}{(kT)^{2}}$. Figure 4 shows the  shear scattering rate
$1/\tau_{2,ee}$ in a temperature range $10 K < T < 40 K$. Fitting the theory to this data yields the following values: $1/\tau_{2,imp}=2\times10^{12}
s^{-1}$ and  C=1. Note, however that fitting theory to experiment in this sample is considerably less accurate
because of a weaker temperature dependence.

In the main text we focused on the results at high density, where there are limits imposed by the hydrodynamic conditions. We indeed analysed
the data for lower voltages, these conditions are no longer being met very well. The figure 5 shows the inverse shear relaxation time for the 3.5 $\mu m$ sample
for  $E_{F}=71 meV$ and comparison with the theory. We fit this data with the following values: $1/\tau_{2,imp}=1.5\times10^{12}
s^{-1}$ and  C=0.7. Note however, that in this case $l_{tr}/l_{2} = 3$, and $l_{2}/W = 3$, which does not completely satisfy all the requirements.
 We are also limited in the use of high gate voltages, because of the possibility of electrical breakdown of the dielectric.

In summary, the dependences R(B) in a wider sample becomes less sensitive to the temperature and magnetic field in accordance with the prediction of the
model \cite{alekseev}. However, we have been able to fit the R(B) curves with the theory which was found to describe the data with a
sufficient accuracy.

\end{document}